%% file: main.tex
\documentclass[10pt,twocolumn,letterpaper]{article}

\usepackage[pagenumbers]{iccv}              

\usepackage{amsmath}
\usepackage{amssymb}
\usepackage{enumitem}
\usepackage{natbib}
\usepackage{graphicx}
\usepackage{float}
\usepackage{physics}
\usepackage{nth}
\usepackage{soul}
\usepackage{xcolor}

\input{preamble}

\definecolor{iccvblue}{rgb}{0.21,0.49,0.74}
\usepackage[pagebackref,breaklinks,colorlinks,allcolors=iccvblue]{hyperref}

\def\paperID{8} 
\def\confName{ICCV}
\def\confYear{2025}

\title{HRVGAN: High Resolution Video Generation using Spatio-Temporal GAN}

\author{Abhinav Sagar\\
University of Maryland, College Park, Maryland\\
College Park, Maryland\\
{\tt\small asagar@umd.edu}
}

\begin{document}

\maketitle

\begin{abstract}

High-resolution video generation has emerged as a crucial task in computer vision, with wide-ranging applications in entertainment, simulation, and data augmentation. However, generating temporally coherent and visually realistic videos remains a significant challenge due to the high dimensionality and complex dynamics of video data. In this paper, we propose a novel deep generative network architecture designed specifically for high-resolution video synthesis. Our approach integrates key concepts from Wasserstein Generative Adversarial Networks (WGANs), enforcing a k-Lipschitz continuity constraint on the discriminator to stabilize training and enhance convergence. We further leverage Conditional GAN (cGAN) techniques by incorporating class labels during both training and inference, enabling class-specific video generation with improved semantic consistency. We provide a detailed layer-wise description of the Generator and Discriminator networks, highlighting architectural design choices promoting temporal coherence and spatial detail. The overall combined architecture, training algorithm, and optimization strategy are thoroughly presented. Our training objective combines a pixel-wise mean squared error loss with an adversarial loss to balance frame-level accuracy and video realism. We validate our approach on benchmark datasets including UCF101, Golf, and Aeroplane, encompassing diverse motion patterns and scene contexts. Quantitative evaluations using Inception Score (IS) and Fréchet Inception Distance (FID) demonstrate that our model significantly outperforms previous state-of-the-art unsupervised video generation methods in terms of both quality and diversity.

\end{abstract}

\section{Introduction}

Deep learning approaches to computer vision have traditionally focused on static image analysis, successfully addressing a wide range of tasks such as classification, detection, and segmentation. However, real-world visual data is inherently dynamic, composed of sequences of frames connected through the temporal dimension. This temporal information carries crucial cues about object motion, scene dynamics, and evolving contexts, enabling richer scene understanding beyond what single images can provide.

Despite this, modeling video data poses unique challenges, primarily due to the increased computational complexity and the need to capture both spatial and temporal dependencies. Static image-based methods are fundamentally limited for tasks such as action recognition and prediction, where understanding motion and temporal progression is essential \cite{huang2018makes}. Consequently, video-based algorithms capable of modeling temporal dynamics have become a critical area of research.

Generating realistic videos from latent representations is a particularly challenging problem in deep learning. State-of-the-art approaches often suffer from blurriness and lack of temporal coherence, highlighting the difficulty of modeling the complex pixel-wise transformations across frames \cite{vondrick2016generating}. Effective video generation requires not only understanding spatial content within each frame but also accurately modeling temporal evolution and uncertainty across frames \cite{villegas2017learning}.

To address these challenges, prior works have sought to disentangle spatial and temporal dynamics. Spatial dynamics capture the appearance and structure of objects in individual frames, while temporal dynamics describe their movement and interactions over time. Methods such as 1-D convolutional networks applied over time \cite{saito2017temporal} and recurrent neural networks (RNNs) generating latent codes for frame synthesis \cite{tulyakov2018mocogan} have been explored to reduce computational cost while modeling temporal dependencies. However, while 1-D convolutions provide efficiency gains, more expressive 3-D convolutional architectures are necessary for accurately modeling the complex spatiotemporal correlations inherent in high-quality video synthesis.

A notable limitation of many existing video generation models is their narrow focus on specialized tasks or datasets, which hinders their generalizability across different video generation and prediction scenarios. Furthermore, most architectures in the literature are tailored to domain-specific problems, lacking the flexibility to adapt to broader settings without significant redesign.

In this work, we propose a novel unsupervised generative adversarial network (GAN) architecture designed for high-resolution video generation and prediction. Our approach addresses these limitations by integrating advanced techniques from Wasserstein GANs and conditional GANs, enabling stable training and improved control over generated content. Crucially, our architecture is designed to be generalizable across diverse datasets and tasks, paving the way for more flexible and scalable video generation frameworks. 

\section{Related Work}

Generative models have shown significant promise in modeling the complex temporal dynamics inherent in video data. Among these, autoregressive models such as those introduced in \cite{van2016conditional} sequentially generate frames conditioned on previous ones, capturing temporal dependencies explicitly. Meanwhile, Generative Adversarial Networks (GANs) have gained traction for video generation due to their ability to produce sharp, realistic frames. Notable GAN-based video generation efforts include \cite{xiong2018learning}, \cite{acharya2018towards}, and \cite{tulyakov2018mocogan}, each demonstrating varying levels of success in balancing temporal coherence and image fidelity.

The progress of GANs in recent years can largely be attributed to improvements in training stability \cite{salimans2016improved}, the introduction of more effective loss functions \cite{deshpande2018generative}, and advances in architectural design such as progressive growing of GANs \cite{karras2017progressive}. However, despite these advances, applying GANs effectively to video data—especially in the context of action prediction and unsupervised settings—remains relatively underexplored.

A core challenge in GAN training is mode collapse, where the generator produces limited diversity in samples, often generating near-identical outputs. To address this, various strategies have been proposed: multi-generator frameworks that encourage diversity, as in \cite{ghosh2018multi}, and reconstructor networks that invert generated samples back to latent codes to enforce consistency \cite{srivastava2017veegan}. Additionally, the progressive growing technique introduced by \cite{karras2017progressive} stabilizes training for high-resolution image synthesis by gradually increasing the network’s capacity and output resolution. Concurrently, improvements in loss formulations such as Wasserstein GAN with Gradient Penalty (WGAN-GP) \cite{arjovsky2017wasserstein, gulrajani2017improved} and perceptual losses \cite{deshpande2018generative} have significantly enhanced training dynamics and output quality.

Much of the existing literature on video generation relies on supervised learning paradigms. For example, \cite{vondrick2016generating} proposed an unsupervised video generation approach that separates foreground and background by employing two parallel streams in the generator—combining 2D and 3D convolutional layers—and a 3D convolutional discriminator. Similarly, \cite{tulyakov2018mocogan} introduced a temporal and spatial generator decomposition, but the output resolution was limited to 64$\times$64 pixels. Another approach by \cite{saito2017temporal} employed a cascade architecture where a temporal generator composed of 1-D deconvolutional layers transforms a latent vector into a sequence of latent vectors, each fed into an image generator to synthesize frames.

Despite these innovations, most existing methods generate low-resolution videos and are often specialized to particular datasets or problem settings, limiting their general applicability.

In contrast, our work advances the state of the art with the following key contributions:

\begin{itemize}
    
\item We propose a novel GAN architecture for unsupervised high-resolution video generation at 256$\times$256 pixels, significantly improving spatial detail and temporal coherence over prior work.

\item We provide detailed architectural descriptions of both the generator and discriminator networks, along with comprehensive explanations of our optimization strategies and composite loss functions combining adversarial and pixel-level objectives.

\item We rigorously validate our approach on publicly available benchmark datasets—including UCF101, Golf, and Aeroplane—demonstrating strong qualitative and quantitative performance.

\item Using standard metrics such as Inception Score (IS) and Fréchet Inception Distance (FID), our network consistently outperforms previous state-of-the-art methods in unsupervised video generation.

\end{itemize}

\section{Background}

\subsection{GAN}

GANs are a family of unsupervised generative models that learns to generate samples from a given distribution \cite{goodfellow2014generative}. Given a noise distribution, Generator $G$ tries to generate samples while the Discriminator $D$ tries to tell whether the generated samples are from the correct distribution or not. Both the generator and discriminator are trying to fool each other, thus playing a zero-sum game. In other words, both are in a state of Nash Equilibrium. Let $G$ represent the generator and $D$ the discriminator, the loss function used for training GAN can be written as shown below:

\begin{align}
\mathcal{F}(\mathcal{D}, \mathcal{G}) =\; & \mathbb{E}_{\mathbf{x} \sim p_{\mathbf{x}}}[-\log \mathcal{D}(\mathbf{x})] \\
& + \mathbb{E}_{\mathbf{z} \sim p_{\mathbf{z}}}[-\log (1-\mathcal{D}(\mathcal{G}(\mathbf{z})))]
\end{align}

where $z$ is latent vector, $x$ is data sample, $p_{z}$ is probability distribution over latent space and $p_{x}$ is
probability distribution over data samples. The zero-sum condition is defined as:

\begin{equation}
\min _{G} \max _{D} \mathcal{F}(\mathcal{D}, \mathcal{G})
\end{equation}

A lot of changes have been proposed over the years to reduce mode collapse and minimize training instability, which are two of the main challenges while training GANs. Some of these changes are using least square loss instead of sigmoid cross entropy loss as shown in \cite{mao2017least} and using feature matching and minibatch discrimination as shown in \cite{salimans2016improved}.

\subsection{Wasserstein GAN}

A new technique was proposed to minimize the Wasserstein Distance (WD) between the distributions to stabilize training. WD between two distributions was used in \cite{arjovsky2017wasserstein} is defined in:

\begin{equation}
W\left(p_{r}, p_{g}\right)=\inf _{\gamma \in \prod\left(p_{r}, p_{g}\right)} \mathbb{E}_{(x, y) \sim \gamma}[\|x-y\|]
\end{equation}

where $p_{r}$, $p_{g}$ are distributions of real and generated samples and $Q(p_{r}, p_{g})$ is the space of all possible joint probability distributions of $p_{r}$ and $p_{g}$.

Another technique, known as weight clipping, was also proposed to enforce the K-Lipschitz constraint. The loss function for training the network is defined as shown in:

\begin{align}
\mathcal{F}(\mathcal{D}, \mathcal{G}) =\;&
\mathbb{E}_{\mathbf{x} \sim p_{\mathbf{x}}}[\mathcal{D}(\mathbf{x})]
- \mathbb{E}_{\mathbf{z} \sim p_{\mathbf{u}}}[\mathcal{D}(\mathcal{G}(\mathbf{z}))] \\
& + \lambda \, \mathbb{E}_{\hat{\mathbf{x}} \sim p_{\mathbf{x}}}
\left[
\left(
\left\| \nabla_{\hat{\mathbf{x}}} \mathcal{D}(\hat{\mathbf{x}}) \right\|_2 - 1
\right)^2
\right]
\end{align}

Where $\lambda$ is a regularization parameter.

\subsection{Conditional GANs}

These types of GANs use conditions on the generator to generate samples with desired properties as first shown in \cite{mirza2014conditional}. The loss functions for Conditional GANs can be defined as in:

\begin{equation}
\mathcal{F}(\mathcal{D}, \mathcal{G})=\mathbb{E}_{\mathbf{x} \sim p_{\times}}[-\log \mathcal{D}(\mathbf{x})]+\mathbb{E}_{\mathbf{z} \sim p_{\mathbf{x}}}[-\log (1-\mathcal{D}(\mathcal{G}(\mathbf{z})))]
\end{equation}

The conditions could be class labels or the original data sample in the case of video prediction.

\section{Method}

\subsection{Dataset}

The following datasets were used in this work for training and testing our network for video generation:

1. UCF101 Dataset: The purpose of this dataset was to train networks robust for action recognition tasks. It contains 13320 videos of 101 different action categories like Sky Diving, Knitting, and Baseball Pitch \cite{soomro2012ucf101}.

2. Golf and Aeroplane Datasets: It contains 128$\times$128 resolution frames, which can be used for evaluating
video generative adversarial networks \cite{vondrick2016generating} and \cite{kratzwald2017towards}.

\subsection{Network Architecture}

Let input sequence frames of a video be denoted by $(X = {X_{1}, ..., X_{m}})$ and frames to be predicted in sequence by $(Y = {Y_{1}, ..., Y_{n}})$. Our proposed network for high-resolution video generation operates in two key stages: (1) a novel conditional Generative Adversarial Network (GAN) designed to generate coherent video sequences conditioned on specified action categories, and (2) a reconstruction network equipped with a newly formulated loss function to refine and map generated latent sequences into the pixel space with high fidelity.

In the first stage, the input to the Generator is a sequence of latent noise vectors, each corresponding to a frame in the target video sequence. These latent vectors are conditioned on the action category label, enabling the Generator to produce temporally coherent frame sequences that semantically align with the specified action class. The Generator transforms the input latent sequence into a sequence of synthesized video frames.

The generated video frames are then passed to the Discriminator, which evaluates the authenticity of the sequence by distinguishing between real video samples from the training dataset and the synthetic frames produced by the Generator. This adversarial setup encourages the Generator to produce realistic and temporally consistent videos.

Both Generator and Discriminator networks are trained end-to-end using mini-batch Stochastic Gradient Descent (SGD) with carefully designed loss functions. The Generator loss combines adversarial feedback with a pixel-level reconstruction loss, ensuring both global realism and local detail preservation, while the Discriminator is trained to robustly classify real versus generated frame sequences.

Architecturally, the Generator leverages 3D transposed convolutional (deconvolutional) layers to jointly model spatial and temporal information, allowing for effective synthesis of video frames that capture both appearance and motion. Conversely, the Discriminator employs 3D convolutional layers to analyze the spatiotemporal coherence of input video sequences. To promote stable and efficient training, batch normalization is applied throughout the Generator network, facilitating gradient flow and convergence, whereas instance normalization is utilized in the Discriminator to improve generalization and reduce internal covariate shift.

Regarding activation functions, the Generator uses ReLU nonlinearities to maintain strong gradient signals during training and encourage sparse feature activations, while the Discriminator employs leaky ReLU activations to mitigate the dying ReLU problem and ensure gradient flow even when activations fall below zero.

The network architecture used in this work is shown in \autoref{figure1}:

\begin{figure}[htp]
    \centering
    \includegraphics[width=8cm]{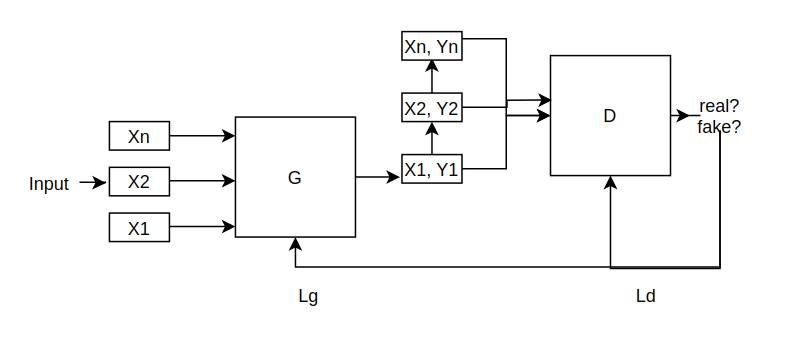}
    \caption{Illustration of the network architecture.}
    \label{figure1}
\end{figure}

The Generator layer-wise details are shown in \autoref{table1}:

\begin{table}[hbt!]
  \caption{Generator architecture layer-wise details.}
  \label{table1}
  \centering
  \begin{tabular}{lll}
  \toprule
    Generator &Activation &Output shape           \\
    \midrule
Latent vector &- &128$\times$1$\times$1$\times$1\\ 
Fully-connected &ReLU &128$\times$1$\times$1$\times$1 \\
DeConv 3$\times$3$\times$3 &ReLU &128$\times$4$\times$4$\times$4 \\
Upsample &- &128$\times$8$\times$8$\times$8 \\
DeConv 3$\times$3$\times$3 &ReLU &128$\times$8$\times$8$\times$8\\\ 
DeConv 3$\times$3$\times$3 &ReLU &128$\times$8$\times$8$\times$8 \\
Upsample &- &128$\times$8$\times$16$\times$16 \\\
DeConv 3$\times$3$\times$3 &ReLU &128$\times$8$\times$16$\times$16 \\
DeConv 3$\times$3$\times$3 &ReLU &128$\times$8$\times$16$\times$16\\\
Upsample &- &128$\times$8$\times$32$\times$32 \\
DeConv 3$\times$3$\times$3 &ReLU &64$\times$8$\times$32$\times$32 \\
DeConv 3$\times$3$\times$3 &ReLU &64$\times$8$\times$32$\times$32 \\
Upsample &- &64$\times$16$\times$64$\times$64 \\
DeConv 3$\times$3$\times$3 &ReLU &32$\times$16$\times$64$\times$64 \\
DeConv 3$\times$3$\times$3 &ReLU &32$\times$16$\times$64$\times$64 \\
Upsample &- &32$\times$16$\times$128$\times$128 \\
DeConv 3$\times$3$\times$3 &ReLU &16$\times$16$\times$128$\times$128\\\
DeConv 3$\times$3$\times$3 &ReLU &16$\times$16$\times$128$\times$128 \\
Upsample &- &16$\times$32$\times$256$\times$256 \\
DeConv 3$\times$3$\times$3 &ReLU &8$\times$32$\times$256$\times$256 \\
DeConv 3$\times$3$\times$3 &ReLU &8$\times$32$\times$256$\times$256 \\
DeConv 1$\times$1$\times$1 &ReLU &3$\times$32$\times$256$\times$256 \\
    \bottomrule
  \end{tabular}
\end{table}

The Discriminator layer-wise details are shown in \autoref{table2}:

\begin{table}[hbt!]
  \caption{Discriminator architecture for generation of 256$\times$256$\times$32 videos}
  \label{table2}
  \centering
  \begin{tabular}{lll}
  \toprule
    Discriminator &Activation &Output shape           \\
    \midrule
Input Image &- &128$\times$1$\times$1\\
Conv 1$\times$1$\times$1 &Leaky ReLU &128$\times$4$\times$4$\times$4 \\
Conv 3$\times$3$\times$3 &Leaky ReLU &128$\times$4$\times$4$\times$4\\\
Conv 3$\times$3$\times$3 &Leaky ReLU &128$\times$4$\times$4$\times$4 \\
Downsample &- &128$\times$8$\times$8$\times$8 \\
Conv 3$\times$3$\times$3 &Leaky ReLU &128$\times$8$\times$8$\times$8 \\
Conv 3$\times$3$\times$3 &Leaky ReLU &128$\times$8$\times$8$\times$8 \\
Downsample &- &128$\times$8$\times$16$\times$16 \\
Conv 3$\times$3$\times$3 &Leaky ReLU &128$\times$8$\times$16$\times$16 \\
Conv 3$\times$3$\times$3 &Leaky ReLU &128$\times$8$\times$16$\times$16\\
Downsample &- &128$\times$8$\times$32$\times$32 \\
Conv 3$\times$3$\times$3 &Leaky ReLU &64$\times$8$\times$32$\times$32 \\
Conv 3$\times$3$\times$3 &LReLU &64$\times$8$\times$32$\times$32\\ 
Downsample &- &64$\times$16$\times$64$\times$64\\
Conv 3$\times$3$\times$3 &Leaky ReLU &32$\times$16$\times$64$\times$64 \\
Conv 3$\times$3$\times$3 &Leaky ReLU &32$\times$16$\times$64$\times$64 \\
Downsample &- &32$\times$16$\times$128$\times$128 \\
Conv 3$\times$3$\times$3 &Leaky ReLU &16$\times$16$\times$128$\times$128 \\
Conv 3$\times$3$\times$3 &Leaky ReLU &16$\times$16$\times$128$\times$128\\ 
Downsample &- &16$\times$32$\times$256$\times$256 \\
Minibatch Stddev &- &129$\times$4$\times$4$\times$4 \\
Conv 3$\times$3$\times$3 &Leaky ReLU &8$\times$32$\times$256$\times$256 \\
Fully-connected &linear &1$\times$1$\times$1$\times$128 \\
Fully-connected &linear &1$\times$1$\times$1$\times$1 \\
    \bottomrule
  \end{tabular}
\end{table}

\subsection{Pixel Normalization}

To avoid an explosion of parameters in both
the generator and the discriminator, feature vectors are normalized at every pixel. We extended the feature vector normalization as proposed by \cite{karras2017progressive} to our spatio-temporal problem.

Let $a_{x,y,t}$ and $b_{x,y,t}$ be the original and normalized feature vectors at pixel $(x, y, t)$ corresponding to spatial and temporal position. The following relation can be written as shown below:

\begin{equation}
b_{x, y, t}=\frac{a_{x, y, t}}{\sqrt{\frac{1}{N} \sum_{j=0}^{N-1}\left(a_{x, y, t}^{j}\right)^{2}+\epsilon}}
\end{equation}

where $\epsilon$ is a constant and $N$ is number of feature maps used.

\subsection{Instance Normalization}

Instance normalization was used after both 3D convolutional and 3D deconvolutional layers to solve the vanishing gradient problem as defined below:

\begin{equation}y=\operatorname{ReLU}\left(\sum_{i=0}^{d} w_{i} \cdot \operatorname{ReLU}\left(\gamma_{i} \cdot \frac{x_{i}-\mu_{i}}{\sqrt{\sigma_{i}^{2}}+\epsilon}+\beta_{i}\right)+b\right)\end{equation}

Where $w$ and $b$ are the weight and bias terms of the 3D convolution layer, $\gamma$ and $\beta$ are the weight and bias terms of the Instance Normalization layer, $\mu$ and $\sigma$ are the mean and variance of the input.

\subsection{Loss Functions}

Generator in the GAN architecture can be used to predict a sequence of frames $Y$ from a sequence of frames $X$ by minimizing the pixel-wise distance between the predicted and the actual frame. The mean square pixel-wise loss function is defined in \autoref{mse_loss}:

\begin{equation}
\mathcal{L}_{mse}(X, Y)=\ell_{mse}(G(X), Y)=\|G(X)-Y\|^{2}
\label{mse_loss}
\end{equation}

The binary cross-entropy loss between the actual and predicted frames is defined in \autoref{bce_loss}:

\begin{equation}
L_{b c e}(Y, \hat{Y})=-\sum_{i} \hat{Y}_{i} \log \left(Y_{i}\right)+\left(1-\hat{Y}_{i}\right) \log \left(1-Y_{i}\right)
\label{bce_loss}
\end{equation}

where both $Y_{i}$ and $Y_{i}$ has values in the range $[0, 1]$.

Let $(X, Y)$ be a sample from the dataset where both $X$ and $Y$ denote a sequence of frames as input and to be predicted, respectively. Let $G$ represent the Generator and $D$ the Discriminator. The goal is to predict the right frames for both the individual classes represented by 0 and 1. The adversarial loss function used for training the Generator is defined in:

\begin{equation}
\mathcal{L}_{a d v}^{G}(X, Y)=\lambda_{1} \sum_{i=1}^{N_{\text {}}} L_{b c e}\left(D_{i}\left(X_{i}, G_{i}\left(X_{i}\right)\right)-k, 1\right)
\end{equation}

The adversarial loss function used for training the Discriminator is defined in:

\begin{align}
\mathcal{L}_{adv}^{D}(X, Y) =\;&
\lambda_{1} \sum_{i=1}^{N} 
L_{bce}\bigl(D_{i}(X_{i}, Y_{i}) - k, 1\bigr) \\
& + \lambda_{2} 
L_{bce}\bigl(D_{i}(X_{i}, G_{i}(X)) - k, 0\bigr)
\end{align}

Where $\lambda_{1}$, $\lambda_{2}$ are the coefficients to balance the penalty terms. $\lambda_{1}$, $\lambda_{2}$ are also used to absorb the scale $k$ caused by the k-Lipschitz constraint on Wasserstein loss.

The mean square loss function and adversarial loss function of the Generator can be combined with equal weights given to both terms, as shown below:

\begin{equation}
\mathcal{L}(X, Y)=\alpha \mathcal{L}_{a d v}^{G}(X, Y)+\beta \mathcal{L}_{mse}(X, Y)
\end{equation}

Where $\alpha$ and $\beta$ are constants with values of 0.5 and 0.5, respectively.

\subsection{Algorithm}

The complete algorithm used in this work is shown below:

\begin{algorithm}
\SetAlgoLined
 Initialize learning rates $\alpha_{D}$ and $\alpha_{G},$ and weights $\lambda_{adv}, \lambda_{\ell_{mse}}$
 
 \While{not converged}{
 
  \textbf{Update the Discriminator} $D$:
  
  Get $M$ data samples $(X, Y)=\left(X^{(1)}, Y^{(1)}\right), \ldots,\left(X^{(M)}, Y^{(M)}\right)$
  
  \[
  W_{D} = W_{D} - \alpha_{D} \sum_{i=1}^{M} \frac{\partial \mathcal{L}_{adv}^{D}\left(X^{(i)}, Y^{(i)}\right)}{\partial W_{D}}
  \]
  
  \textbf{Update the Generator} $G$:
  
  Get $M$ data samples $(X, Y)=\left(X^{(1)}, Y^{(1)}\right), \ldots,\left(X^{(M)}, Y^{(M)}\right)$
  
  \[
  \begin{aligned}
  W_{G} = W_{G} - \alpha_{G} \sum_{i=1}^{M} \Biggl(
  & \lambda_{adv} \frac{\partial \mathcal{L}_{adv}^{G}\left(X^{(i)}, Y^{(i)}\right)}{\partial W_{G}} \\
  & + \lambda_{\ell_{mse}} \frac{\partial \mathcal{L}_{\ell_{mse}}\left(X^{(i)}, Y^{(i)}\right)}{\partial W_{G}}
  \Biggr)
  \end{aligned}
  \]

 }
 \caption{HRVGAN: High Resolution Video Generation using Spatio-Temporal GAN}
\end{algorithm}

\subsection{Evaluation Metrics}

Various metrics have been proposed for evaluating GANs in the literature. Two of the most common metrics are Inception Score and Fréchet Inception Distance, which are explained below:

1. \textbf {Inception Score (IS)} - Inception Score was first proposed in \cite{salimans2016improved} for evaluating GANs. A higher inception score is preferred, which means the model can generate diverse images, thus avoiding the mode collapse issue.

Let $x$ be samples generated by the generator $G$, $p(y|x)$ be the distribution of classes for
generated samples, and $p(y)$ be the marginal class distribution. The Inception score is defined as:

\begin{equation}
I S(\mathcal{G})=\exp \left(\mathbb{E}_{\mathbf{x} \sim p_{g}} \mathcal{D}_{K L}(p(y \mid \mathbf{x})|| p(y))\right)
\end{equation}

where $D_{KL}$ is the Kullback-Leibler divergence between $p(y|x)$ and $p(y)$.
 
2. \textbf {Fréchet Inception Distance (FID)} - Another metric to evaluate the quality of generated samples was first proposed by \cite{heusel2017gans}. 

Let $D$ represent the CNN used to extract features, ($m_{r}$, $\sigma_{r}$) be mean and covariance of features extracted from real samples and ($m_{f}$, $\sigma_{f}$) be mean and covariance of features extracted from fake samples with $D$, then the Frechet Inception distance is defined as:

\begin{align}
d^{2}\bigl((m_{r}, \Sigma_{r}), (m_{f}, \Sigma_{f})\bigr) =\;&
\left\| m_{r} - m_{f} \right\|_{2}^{2} \\
& + \operatorname{Tr}\left(
\Sigma_{r} + \Sigma_{f} - 2 \left( \Sigma_{r} \Sigma_{f} \right)^{\frac{1}{2}}
\right)
\end{align}

Fréchet Inception Distance is more accurate than Inception Score as it compares summary statistics of generated samples and real samples. A lower FID is preferred for better-performing generative models. 

\section{Results and Discussion}

The Inception Scores of our proposed model compared to existing approaches on the UCF101 dataset are summarized in \autoref{table3}. Our method achieves a significant improvement over prior state-of-the-art models, indicating superior generative quality and diversity in the synthesized videos.

\begin{table}[hbt!]
\caption{Comparison of Inception Scores on the UCF101 dataset. Higher scores indicate better video quality and diversity.}
\label{table3}
\centering
\begin{tabular}{ll}
\toprule
Model & Inception Score \\
\midrule
VGAN \cite{vondrick2016generating} & 8.18 \\
TGAN \cite{saito2017temporal} & 11.85 \\
MoCoGAN \cite{tulyakov2018mocogan} & 12.42 \\
Ours & \textbf{14.29} \\
\bottomrule
\end{tabular}
\end{table}

We further evaluate our network’s performance using the Fréchet Inception Distance (FID) metric on the Golf and Aeroplane datasets, as shown in \autoref{table4}. The FID score measures the similarity between the distributions of real and generated videos, with lower values indicating better generation quality. Our model consistently outperforms both VGAN and TGAN by achieving substantially lower FID scores across both datasets.

\begin{table}[hbt!]
\caption{Quantitative comparison of FID scores on Golf and Aeroplane datasets. Lower FID scores indicate better video generation quality.}
\label{table4}
\centering
\begin{tabular}{lll}
\toprule
Model & FID Score (Golf) & FID Score (Aeroplane) \\
\midrule
VGAN \cite{vondrick2016generating} & 113,007 & 149,094 \\
TGAN \cite{saito2017temporal} & 112,029 & 120,417 \\
Ours & \textbf{102,584} & \textbf{104,036} \\
\bottomrule
\end{tabular}
\end{table}

\begin{figure}[htp]
\centering
\includegraphics[width=8cm]{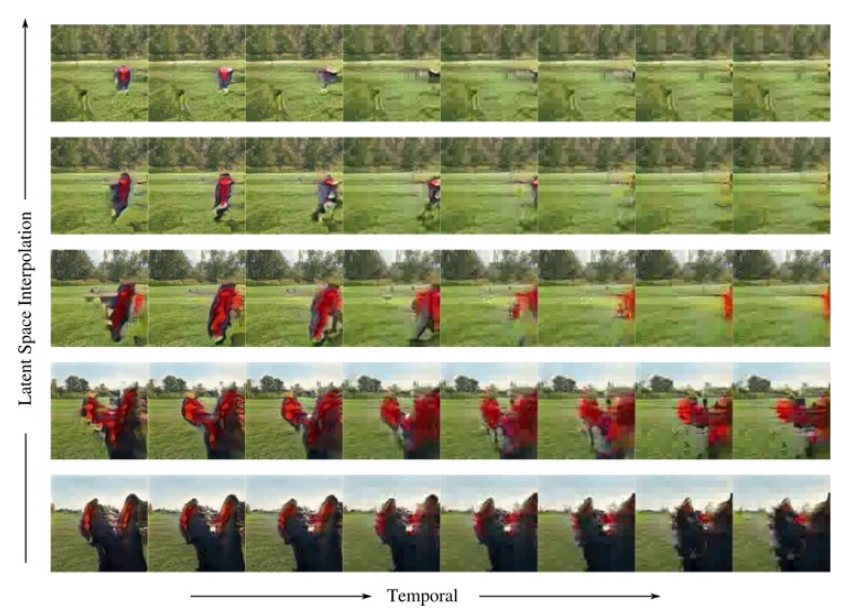}
\caption{Linear interpolation in latent space generating samples from the Golf dataset, showing smooth transitions between video frames.}
\label{fig2}
\end{figure}

To demonstrate the smoothness and semantic consistency of our learned latent space, we perform linear interpolation experiments. \autoref{fig2} illustrates the linear interpolation results on the Golf dataset, where gradual transitions between video samples highlight the model’s ability to capture meaningful latent representations.

\begin{figure}[htp]
\centering
\includegraphics[width=8cm]{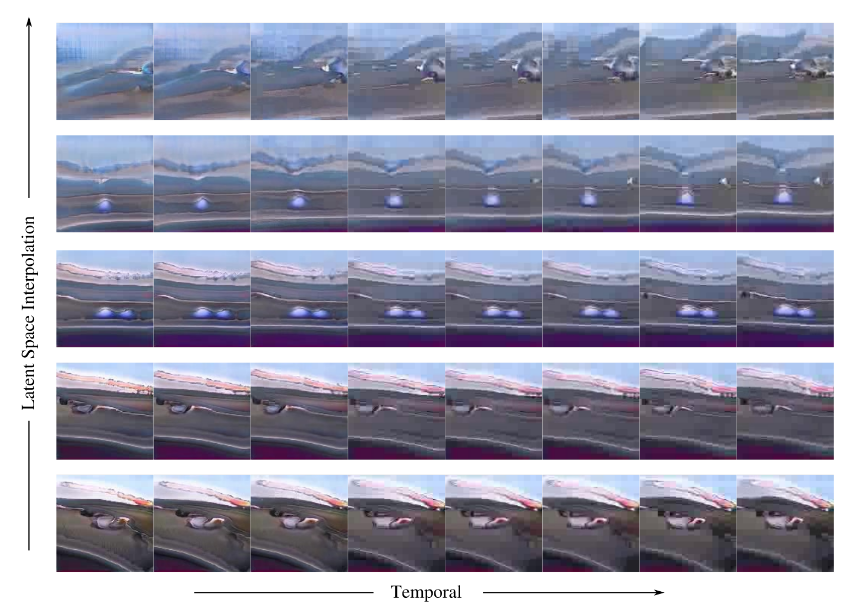}
\caption{Latent space interpolation results on the Aeroplane dataset, illustrating smooth generation between video samples.}
\label{fig3}
\end{figure}

Similarly, \autoref{fig3} presents latent space interpolation on the Aeroplane dataset, further confirming the model’s capacity for generating coherent intermediate video frames across distinct samples.

\begin{figure}[htp]
\centering
\includegraphics[width=8cm]{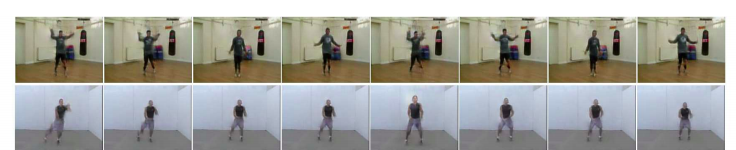}
\caption{Generated frames on UCF101 dataset for JumpingJack (1st row) and TaiChi (2nd row) actions. Each sequence shows 8 frames generated from random latent vectors, highlighting temporal consistency and realism.}
\label{fig4}
\end{figure}

Finally, qualitative results on the UCF101 dataset are shown in \autoref{fig4}. Here, we present generated video frames for two action classes: JumpingJack (top row) and TaiChi (bottom row), each consisting of 8 consecutive frames synthesized from random noise. The results demonstrate the temporal coherence and high visual fidelity of our generated videos.

\section{Conclusions}

In this paper, we propose a novel neural network architecture leveraging generative models for unsupervised video generation. Our approach extends the original GAN framework and is trained using mini-batch stochastic gradient descent. The training objective combines a mean squared pixel loss with an adversarial loss constrained by a k-Lipschitz condition, inspired by Wasserstein GANs. We provide comprehensive details on the network architecture, optimization strategy, and the full algorithmic pipeline. Experimental results on the UCF101, Golf, and Aeroplane datasets demonstrate that our model surpasses existing state-of-the-art methods, evaluated through Inception Score and Fréchet Inception Distance metrics. Additionally, we showcase smooth linear interpolations in the latent space for the Golf and Aeroplane datasets, along with qualitative frame generation results on UCF101.

\nocite{*}
{
    \small
    \bibliographystyle{ieeenat_fullname}
    \bibliography{main}
}

\end{document}

%% file: preamble.tex
\usepackage[ruled,vlined]{algorithm2e} 
